\documentclass[american,copyright]{eptcs}
\providecommand{\event}{MBT 2015}

\usepackage[utf8]{inputenc}
\usepackage{color}
\usepackage{amsmath}
\usepackage{graphicx}
\usepackage{listings}
\usepackage{xspace}
\usepackage{listings}
\usepackage{amsfonts}

\usepackage{color}

\definecolor{gray}{rgb}{0.5,0.5,0.5}
\definecolor{lightgray}{rgb}{0.83, 0.83, 0.83}

\newcommand{\monkey}{{\sc Monkey}\xspace}
\newcommand{\monkeyr}{{\sc MonkeyRunner}\xspace}
\newcommand{\spin}{{\sc spin}\xspace}
\newcommand{\promela}{{\sc promela}\xspace}
\newcommand{\android}{{\sc android}\xspace}
\newcommand{\dragonfly}{{\sc dragonfly}\xspace}
\newcommand{\java}{{\sc java}\xspace}
\newcommand{\xml}{{\sc XML}\xspace}

\newcommand{\uiautomatorviewer}{{\sc UIautomatorViewer}\xspace}
\newcommand{\uiautomator}{{\sc UIautomator}\xspace}

\newcommand{\machine}[1]{\ensuremath{M_{#1} = \langle \Sigma_{#1},I_{#1},\xrightarrow{-}_{#1},E_{#1},C_{#1},F_{#1} \rangle}}

\lstdefinestyle{pseudocode}{basicstyle=\ttfamily\scriptsize}
\lstdefinestyle{xml}{tabsize=2}

\newtheorem{definition}{Definition}

\graphicspath{{Images/}}

\title 
\makeatletter
\def\verbatim@font{\ttfamily\small}
\makeatother

\title{Using Model Checking to Generate Test Cases for Android Applications}
\author{
Ana Rosario Espada \qquad Mar\'ia del Mar Gallardo \\ Alberto Salmer\'on \qquad Pedro Merino
\institute{Dept. Lenguajes y Ciencias de la Computaci\'on\\
E.T.S.I. Inform\'atica \quad
University of M\'alaga\thanks{Work partially supported by grants P11-TIC-07659 (Regional Government of Andalusia), TIN2012-35669 (Spanish Ministry of Economy and Competitiveness), UMA-Agilent Technologies 808/47.3868- 4Green and the AUIP as sponsor of the Scholarship Program Academic Mobility.}}
\email{[anarosario,gallardo,salmeron,pedro]@lcc.uma.es }}

\def\authorrunning{Espada, Gallardo, Salmerón \& Merino}
\def\titlerunning{Using Model Checking to Generate Test Cases for Android Applications}

\begin{document}

\maketitle

\begin{abstract}
The behavior of mobile devices is highly non deterministic and barely predictable due to the interaction of the user with its applications. In consequence, analyzing the correctness of applications running on a smartphone involves dealing with the complexity of its environment. In this paper, we propose the use of \emph{model-based testing} to describe the potential behaviors of users interacting with mobile applications. These behaviors are modeled by composing specially-designed state machines. These composed state machines can be exhaustively explored using a model checking tool to automatically generate all possible user interactions. Each generated trace model checker can be interpreted as a test case to drive a runtime analysis of actual applications. We have implemented a tool that follows the proposed methodology to analyze \android\ devices using the model checker \spin\ as the exhaustive generator of test cases.



\end{abstract}



\section{Introduction}
\label{sec:introduction}

At present, smartphone technology is ubiquitous and changes
constantly. Users use their mobiles not only as phones, but as
compact computers, able to concurrently provide services which are
rapidly created, updated, renewed and distributed. In this scenario
of continuous evolution, different
operating systems have been developed such as {\sc symbiam, ios, windows phone} and {\sc  \android},
which allow phones to  support more and more
complex applications.
These platforms define new models of execution, quite different from those used by non-mobile devices.
For instance, one of the most defining characteristics of these systems is their open and event-driven nature. Mobile devices execute a continuous cycle that consists of  first, waiting for the user input and second, producing a  response according to that input. In addition, the internal structure of mobile systems is constructed from a complex combination of
applications,  which enable users to easily navigate through  them.
Thus, although, at a lower level, the execution of applications on a mobile device involves the concurrent execution of several processes (for instance, in \android, applications are \java\ processes executing on the underlying  {\sc linux}  operating system), the way these
applications interact with each other and with the environment does not correspond with the standard interleaving based concurrency model.


It is clear that the execution of  applications on these new operating systems, such as \android~\cite{AndroidDev}, may lead to the
appearance of undesirable bugs which may cause
the phone to malfunction. For example, mobile devices may display the typical errors of
concurrent systems such as violations of safety and liveness
properties. However, there are other
bugs inherent to the particular concurrency model supported by
the new platforms which are not directly analyzable using current
verification technologies.
For example, applications could incorrectly implement the
life cycles of their activities or services (in the case of \android), or may misbehave upon
the arrival of unexpected  external events. In addition, conversion errors, unhandled exceptions, errors of
incompatibility {\sc api}  and {\sc i/o} interaction errors as
described in \cite{Hu} may also appear. 

Different techniques for analyzing the execution of mobile platforms have been proposed. Verification approaches such as model checking~\cite{CGP1999} can be  applied to the software  for mobile devices ~\cite{jpf,jpf2,LX2013}. Model checking is based on a exhaustive generation of all the inter leavings for the threads/processes. A major problem to apply this technique to the real code, like mobile applications, it the need to construct a model of the underlaying operating system or libraries ~\cite{SocketMC,ArincTester,Dwyer04}.  The open nature of these platforms, which are continuously interacting with an unspecified environment,  makes other analysis techniques such as  {\em testing}, {\em monitoring}, and {\em runtime verification} more suitable to check bugs without too much extra effort to model the operating system or the libraries. There have been several recent proposals~\cite{EGCC2010,TKH2011,KFBJB2014} for testing in this framework with commercial tools~\cite{Robotium,DroidPilot}. In these approaches, test cases are randomly generated with tools such as \monkey and \monkeyr~\cite{AndroidDev}.

Testing and runtime verification maybe also combined, as described in \cite{ABGH2005}, to construct verification tools for mobile applications~\cite{MTN2013,YCZJ2013}. On the one hand, the careful selection of test cases guides the execution of the device, while, on the other, the runtime verification module implements observers devoted to analyzing the traces produced by the device. The runtime verification module was already addressed by some authors of this paper in~\cite{Drangofly}. Here we focus on describing how the generation of test cases may be carried out following the  {\em model-based testing} approach~\cite{UL2006} supported by model checking tools.


Our proposal is based on the idea that although the interaction between the user and the mobile device is completely undetermined, each application is associated with a set of {\em intended user behaviors} which define the {\em common ways} of using the application. For each application, or more precisely, for each application view, we use state machines to  construct a {\em non deterministic model} representing the expected use of the view/application. This state machine is built semiautomatically, with information provided by the expert (the app designer or tester) and by \android\ supporting tools like \uiautomatorviewer. Then, all these view models may be conveniently composed to construct a {\em non deterministic} model of the user interaction with a subset of mobile applications of interest. Due to the way of building the state machines, each execution trace of the {\em composed state machine} corresponds to a possible realistic use of the mobile. Thus, the generation of test cases is reduced to the generation of all possible behaviors of the composed machine, which may be carried out by a model checking tool. Although the methodology proposed does not depend on the underlying mobile operating system, the tool has been built on the assumption that the operating system is \android. 

The paper provides two main contributions. The first one is  the formal definition of a special type of state machine that models the expected user interaction with the mobile application.  The approach to modeling is completely modular in the sense that adding (or removing) new view state machines to incorporate (eliminate) user behaviors does not affect the rest of state machines that have already been defined. The second one is a method to employ the explicit model checker \spin~\cite{holzmann-03} that takes the composed state machine as input and and produces a significant set of test cases that generate traces for runtime verification tools. We have constructed a tool chain which implements both modeling and test generation phases to shows the  feasibility of the approach in practice.


The rest of the paper is organized as follows. Section~\ref{sec:modelchecking} describes our approach to using model checking for test case generation. Section~\ref{sec:overview} introduces the testing platform that we are developing. Section~\ref{sec:formal} provides a formal description of the behaviour of composed state machines. Section~\ref{sec:Implementation} uses well known \android\ applications to describe how our approach for test case generation is implemented.  Section~\ref{sec:related} gives a comparison with related work.  Last section summarizes conclusions and future work.

\section{Model checking for test case generation}
\label{sec:modelchecking}


\spin~\cite{holzmann-03} is a model checker that can be used to verify the correctness of concurrent software systems modeled using the specification language \promela. The focus of the tool is on the design and validation of computer protocols, although it has been applied to other areas. \spin can check the occurrence of a property over all possible executions of a system specification, and provide counterexamples when violations are found.
We use the \spin model checker in our approach for automatically generating test cases from application models in the following way. First, each device will be represented by a single \promela process, which contains a state machine that models the applications contained on that device. The state machines themselves are written as loops, where each branch corresponds to a transition triggered by an event. The current state of each state machine (stored as a global \promela variable) determines which branches are active and may be taken. The right hand side of each branch records the transition and updates the current state. This \promela specification is explored exhaustively by \spin in order to generate all possible test cases described by the application model, taking all possible alternatives when there is more than one active branch at the same time.

Listing~\ref{lst:modelchecking:testcases:promela} shows an example of a \promela specification that follows the approach outlined above. This example contains two devices and with their corresponding state machine (\texttt{device\_A()} in line~\ref{lst:modelchecking:testcases:promela:devicea} and \texttt{device\_B()} in line~\ref{lst:modelchecking:testcases:promela:deviceb}), with two states plus the initial state. The \texttt{transition} function is used to record the user or system transition associated with each branch. In order to complete a test case, all devices must have finished their respective state machines (lines~\ref{lst:modelchecking:testcases:promela:finisha} and~\ref{lst:modelchecking:testcases:promela:finishb}), usually when the \texttt{do} loop is exited (lines~\ref{lst:modelchecking:testcases:promela:breaka} and~\ref{lst:modelchecking:testcases:promela:breakb}). This enables the \texttt{traceCloser} process to be executed due to its schedulability restrictions (line~\ref{lst:modelchecking:testcases:promela:tracecloser}), which prints the transitions of the generated test case.

\begin{lstlisting}[caption={Sample \promela specification for test generation},float=t,label=lst:modelchecking:testcases:promela]
mtype = { state_init, state_1, state_2 };
typedef Device { byte transitions[MAX_TR]; short index; bool finish; } /*@ \label{lst:modelchecking:testcases:promela:maxtr} @*/
Device devices[DEVICES]; /*@ \label{lst:modelchecking:testcases:promela:devicetransitions} @*/
mtype state[DEVICES]; /*@ \label{lst:modelchecking:testcases:promela:state} @*/
active proctype traceCloser() provided (devices[DEVA].finish && devices[DEVB].finish) { /*@ \label{lst:modelchecking:testcases:promela:tracecloser} @*/
end_tc:	outputTransitions()
}
active proctype device_A() { /*@ \label{lst:modelchecking:testcases:promela:devicea} @*/
	state[DEVA] = state_init;
	do
	:: state[DEVA] == state_init -> transition(DEVA, BUTTON_1); state[DEVA] = state_1
	:: state[DEVA] == state_1 -> transition(DEVA, SWIPE); state[DEVA] = state_1 /*@ \label{lst:modelchecking:testcases:promela:statea} @*/
	:: state[DEVA] == state_1 -> transition(DEVA, BUTTON_2); state[DEVA] = state_2 /*@ \label{lst:modelchecking:testcases:promela:stateb} @*/
	:: state[DEVA] == state_2 -> transition(DEVA, MESSAGE); break /*@ \label{lst:modelchecking:testcases:promela:system}\label{lst:modelchecking:testcases:promela:breaka} @*/
	:: state[DEVA] == state_2 -> transition(DEVA, BACK); break /*@ \label{lst:modelchecking:testcases:promela:breakb} @*/
	od;
	devices[DEVA].finish = true; /*@ \label{lst:modelchecking:testcases:promela:finisha} @*/
}
active proctype device_B() { /*@ \label{lst:modelchecking:testcases:promela:deviceb} @*/
	state[DEVB] = state_init;
	...
	devices[DEVB].finish = true; /*@ \label{lst:modelchecking:testcases:promela:finishb} @*/
}
\end{lstlisting}

In addition to the current state (line~\ref{lst:modelchecking:testcases:promela:state}), this \promela specification also keeps a list of the transitions taken on the test currently being generated (line~\ref{lst:modelchecking:testcases:promela:maxtr}). The purpose of this data structure is twofold. On the one hand, \texttt{outputTransitions} will print the trace stored here. On the other hand, the history of the current trace is kept inside the \spin's global state, which is taken into consideration when deciding if a state has already been visited. Thus, the same transition may be taken more than once if possible (e.g. line~\ref{lst:modelchecking:testcases:promela:statea}), since the history of the states will be different. However, this requires the maximum depth of exploration to be bounded by the \texttt{MAX\_TR} constant (line~\ref{lst:modelchecking:testcases:promela:maxtr}).

\section{Architecture of the platform}
\label{sec:overview}



\begin{figure}[t]
\centering
\includegraphics[width=0.6\columnwidth]{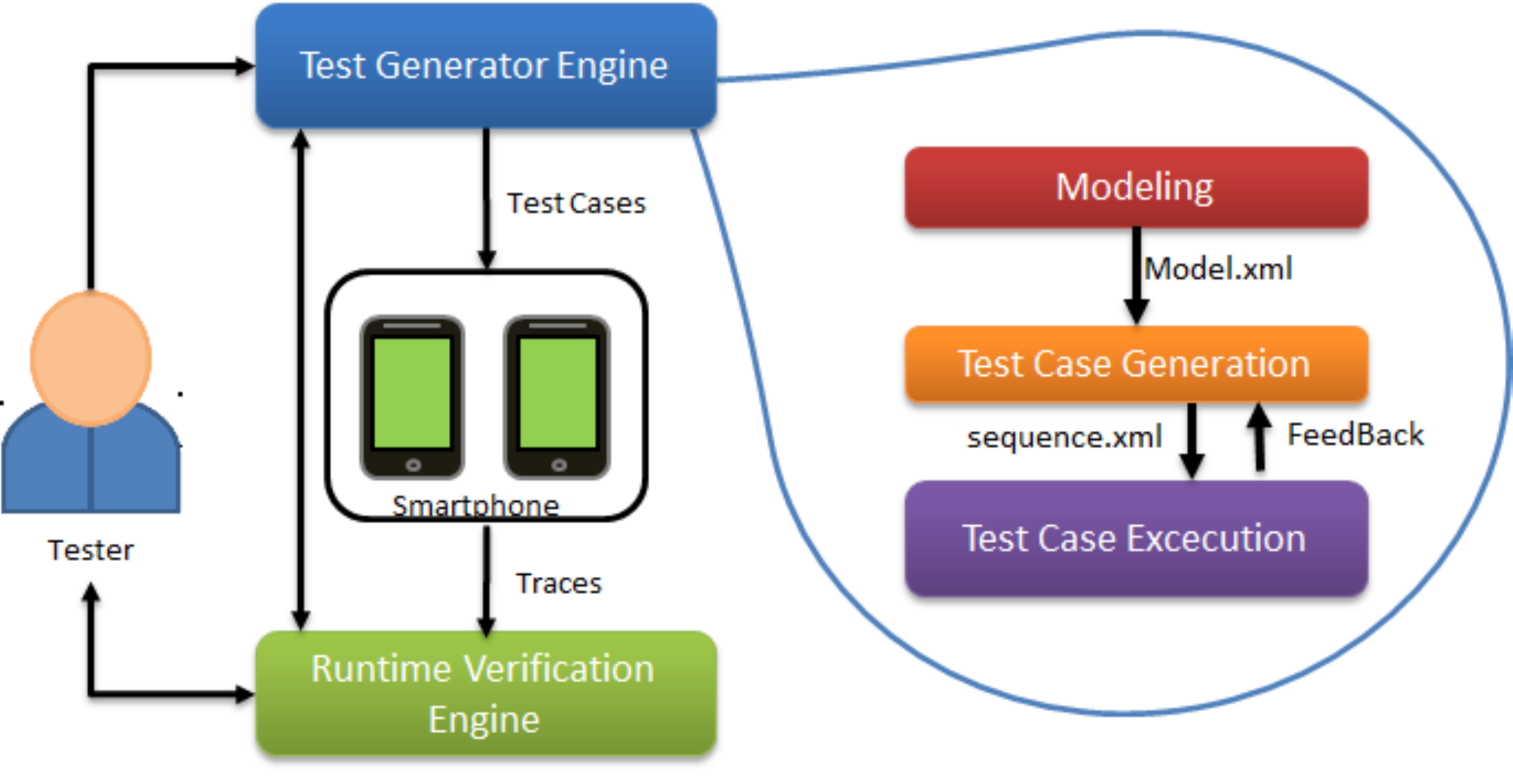}
\caption{Architecture}
    \label{fig:scenario}
\end{figure}

Figure~\ref{fig:scenario} shows the general structure of tools that combine testing and runtime verification techniques to analyze the behavior of applications running on mobile devices. The bottom side uses observers/monitors to analyze the resulting execution traces and verify whether they comply with the expected properties as implemented in the tool \dragonfly~\cite{Drangofly,prole}. The top side of the figure shows the generation of test cases considered in this paper.   The Tester is the expert responsible for modeling the behavior of the applications to be analyzed using a state chart diagram. These models may be constructed as part of the design phase of the applications, and are characterized by their compositional nature: functionality can be added to an existing view without essentially altering the existing behavior.

\begin{figure}[t]
\centering
\includegraphics[width=0.65\columnwidth]{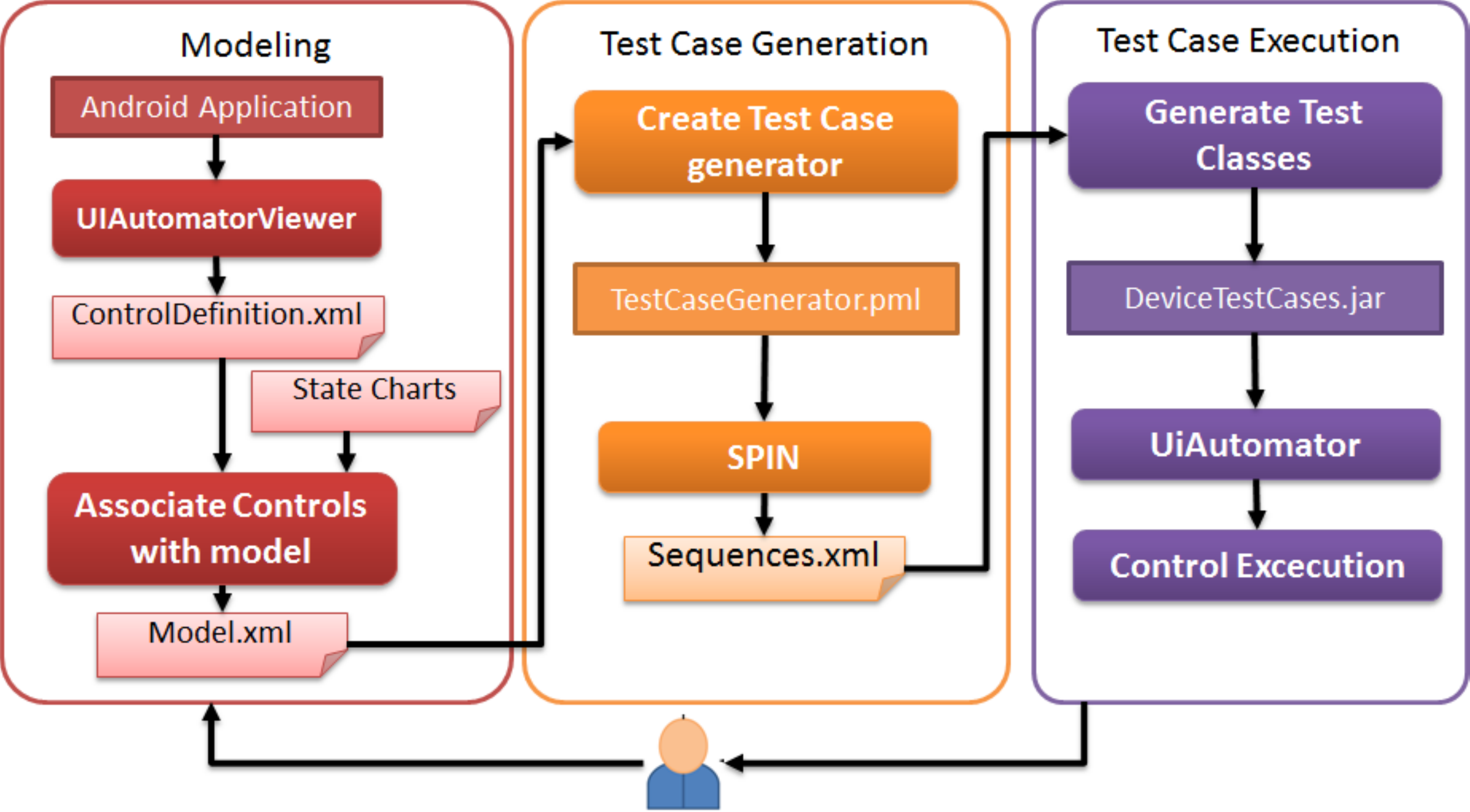}
\caption{Test generation and execution process}
    \label{fig:basedmodeltesting}
\end{figure}

Figure~\ref{fig:basedmodeltesting} shows the complete process of our actual proposal for test generation and execution, which is divided into three main modules:


 \begin{itemize}

\item {\em Modeling.} \uiautomatorviewer tool from \android tools extracts the controls definition in each view of the \android application under analysis. Then, the  controls definition and the state chart diagrams are associated into a \texttt{Model.xml} file with a given structure.

 \item {\em Test Case Generation.} Creates a test case generator per \xml file model into a \promela file.  The \spin model checker~\cite{holzmann-03} performs an exhaustive search of all valid paths in the model using the method explained in Section~\ref{sec:modelchecking}, which correspond to test cases, and generates an \xml file for each one with the appropriate sequence of user input events.

 \item {\em Test Case Execution.} Generates each test class provided using the valid paths described into \xml file by the test case generating module. Then, they can be executed by the \android framework, and sends them to the devices to be executed using the \uiautomator tool which is an extension of JUnit tool using to write user interfaces test cases for \android.
%
\end{itemize}

The following sections provide details about the internal behavior of the Modeling and  Test Case Generator modules, which are the aim of this work. 

\section{Formal Description of models}
\label{sec:formal}

In the following description, we define the behaviour of mobile applications through the composition of state machines at  different abstraction levels. The lowest level is composed of {\em view state machines}. A view corresponds to a mobile screen, with its buttons, text fields, etc.. through which users may interact with the device. When the view is active, users may fire  events through its interface.  A {\em view state machine} models the possible  behaviors of the user when he/she is making use of the view. These {\em behaviors} coincide with the sequence of events fired by the user. Sometimes one of these events  makes a different view becomes active. We have modeled this control transfer between views through the {\em composition relation} of view state machines  from which  {\em device state machines} are constructed. Device state machines  use the {\em connection states} to switch from the current active view to a different view. In this formalization, the specific applications to which each view belongs have not been taken into account, that is, we only model the transfer from one view to another, irrespective of whether both views belong to the same application. 
In the sequel, we use symbols $\xrightarrow{-}/\xrightarrow{-}_i$ to denote the transition relation of the view state machines $M/M_i$. In addition, symbol $\xrightarrow{-}_c$ defines the binary relation which allows us to connect view state machines. Finally,  $\xrightarrow{-}_d$ represents the transition relation of the device state machine which is constructed from
relations $\xrightarrow{-}/\xrightarrow{-}_i$ and $\xrightarrow{-}_c$.

Since \android\ applications are event driven, we may consider that each {\em test case} corresponds to the sequence of  events fired which drive the mobile behaviour. In the formal description, events are the labels of  transitions ($\xrightarrow{-}/\xrightarrow{-}_i$, $\xrightarrow{-}_c$, $\xrightarrow{-}_d$) and have the natural meaning. For instance,  $s \xrightarrow{e} s'$ means that event $e$ must be fired to be able to transit from $s$ to $s'$.

\subsection{View state machines}


\begin{definition}
A {\em view state machine} is a tuple  $M=\langle \Sigma,I,\xrightarrow{-}~,E,C,F\rangle$, where $\Sigma$ is a finite set of states, $I \subseteq \Sigma$ are the initial states,
$C \subseteq \Sigma$ are the so-called {\em connection states}, $F\subseteq \Sigma$ is the set of final states, $E$ is the set of user events,
and $\xrightarrow{-} \subseteq \Sigma  \times E \times \Sigma$ is the labelled transition relation. Sets $I$, $C$ and $F$ are mutually disjoint.
\end{definition}
{\em Final states} are states from which it is not possible to evolve. {\em Connection states} are states from which it is possible to transit a different state machine. These states are essential to model the switch between typical views of smart phone devices. Usually, when a new view is called, the execution of the system is supposed to return to the view caller. To take this behavior into account, we assume that each connection state $s \in C$ has a related state $return(s) \in \Sigma$ which represents the state to be returned when the new view invoked from $s$ has finished its execution.

 We partition  set of events $E$ into two disjunct sets: the set of {\em user events}, denoted as $E^+$, which contains events such  pressing a button, etc., and the set of {\em system events}, denoted as $E^-$, which includes, for instance, events corresponding to system responses to user requests. In the following, we use  $e^+$, $e^-$  to represent user events and system events, and we use $e$ to refer to events which may be of any of both types.

View state machines are deterministic in the sense that if $s \xrightarrow{e} s_1$, and $s \xrightarrow{e'} s_2$ and $e = e'$, then $s_1 = s_2$ that is,  the machine defines, at most, a transition for each pair state/input event.

We now define the notion of flow (an execution in a view state machine), and the test cases generated from flows.
 \begin{definition}\label{lab:flow-test-cases}
 Given a view state machine  $M = \langle \Sigma,I,\xrightarrow{-},E,C,F \rangle$,
 we define the set $Flow(M) = \{s_0 \xrightarrow{e_1} s_1\xrightarrow{e_2} \cdots \xrightarrow{e_{n}} s_n | s_0 \in I, s_n \in F \cup C\}$ of all sequences of states, allowed by $M$, starting at an initial state of $M$, and ending at a final or connection state of $M$.
 The {\em length} of a flow is the number of its states. Given a flow of length $n$, $\phi =  s_0 \xrightarrow{e_1}  \cdots  \xrightarrow{e_n} s_n\in Flow(M)$, the sequence of  events determined by $\phi$ (the test case) is $test(\phi)=e_{1}\cdot \cdots \cdot e_{n}$. We define the set of test cases allowed by $M$ as $TC(M)=\{test(\phi) | \phi \in Flow(M)\}$.
\end{definition}

According with Definition~\ref{lab:flow-test-cases}, test cases are finite sequences of user and system events. For instance, sequence $e^+_1\cdot e^+_2 \cdot e^-_3 \cdot e^+_4$  represents a test case where the user first fires events $e^+_1$ and $e^+_2$, then the system fires $e^-_3$, and finally user fires $e^+_4$. Thus, user and system events are similarly dealt with during the generation of test cases. The difference between them is of importance when  test cases are transformed into code to be executed on the mobile as described in Section~\ref{sec:casestudy}. User events will be transformed into non-blocking  calls to methods that simulate the real occurrence of the event, while system events will correspond to calls to blocking methods which wait for the arrival of the system event.




\subsection{Composition of view state machines}

In this section, we describe how view state machines are composed to construct flows that navigate through different views representing  realistic ways of using a mobile.

We first define the transition between different view state machines. This transition is realized through the {\em binary relation} $\mathcal{R}$ defined between the connection and initial states. The idea is as follows. Assume that the flow in execution belongs to a view state machine $M_i$, and that a connection state $cs$ of $M_i$ has been reached. If  relation $\mathcal{R}$  defines a  transition from $cs$ to some initial state of other machine $M_j$, the flow could jump from $M_i$ to $M_j$, and proceed following the transition relation of $M_j$.
This jump implies  the change in the activity visible in the device from $M_i$ to $M_j$.  
 In the sequel, we call {\em active} the view state machine which is visible in the device, and {\em create} to the rest of view state machines which have been created but are not currently visible in the device.

Given a finite family of state machines
 $\machine{i}$
  we define
 $\Sigma =  \cup_{i=1}^n\Sigma_i$, $I = \cup_{i=1}^nI_i$, $E = \cup_{i=1}^nE_i$, $C = \cup_{i=1}^nC_i$, and $F = \cup_{i=1}^nF_i$. In addition, we denote with $ \mathcal{E} \subseteq E$ the set of {\em call events} that provoke the switch between active view state machines.

 \begin{definition}
Let us assume  a finite family of state machines, \machine{i}.   The {\em connection of view state machines} $M_1,\cdots, M_n$ is given by a binary relation $\mathcal{R}(M_1,\cdots,M_n)\subseteq C \times \mathcal{E} \times I$, that connects connection states with initial states. 
In the following, we denote 3-uples $(s_i,e,s_j)$ of $\mathcal{R}(M_1,\cdots,M_n)$ as $s_i \xrightarrow{e}_c s_j$. Observe that source  and target machines $i$ and $j$ may coincide.



\end{definition}

When a new view is created, the call event may specify some parameters that determine how it must be started or finished. For instance, if the view has already been created, the caller may choose whether to reuse the previously created view or, to the contrary, create a new one. Additionally, when the new created view has finished the execution, the caller view may automatically become active or not. Boolean functions $reuse,auto\_return:\mathcal{E} \to \{\mathit{false},true\}$ establish these parameters for the call events. Although there are other parameters that can be defined in the call events, these two are sufficient to describe the mobile behaviour.





We now define the {\em device state machine} which composes the behavior displayed by the view state machines using the connection relation.

\begin{definition}
Let us assume a finite family of state machines, $M_i = \langle \Sigma_i,I_i,\xrightarrow{-}_i,E_i,C_i,F_i \rangle$, and a connection relation of $M_1,\cdots,M_n$, $\mathcal{R}(M_1,\cdots,M_n)$,  as defined above.
The {\em device state machine} $${\mathcal D}=M_1|||\cdots|||M_n ||| \mathcal{R}(M_1,\cdots,M_n)  $$
 
 \noindent is defined  as the state machine
$\langle \Sigma \times \Sigma^*\times \mathcal{E}^*,I,\xrightarrow{-}_d,E ,F \rangle$ where

\begin{enumerate}
\item $\Sigma^*$ is the set of finite sequences of states of $\Sigma$, and $\mathcal{E}^*$ is the set of finite sequences of call events.

\item Transition relation $\xrightarrow{-}_d$ is defined by the rules below.
\end{enumerate}
\end{definition}

We call {\em configurations} the states of device state machines. A configuration is a 3-uple $\langle s,h,eh \rangle$ where $s$ is the current state of the active view state machine, i.e., the view visible in the mobile. Sequence $h$ is the stack of states $s_1\cdot s_2 \cdots s_n$ which constitute the history of the view machines which have been created (and have not been yet destroyed) in the device but which are not currently visible. Each state $s_i$  of $s_1\cdot s_2 \cdots s_n$ is a connection state of a view state machine which was active, but a transition from $s_i$ to another view machine took place, and the view state machine of $s_i$ became inactive.   Finally $eh= e_1 \cdots e_n$ is the history of events that  have provoked a view switch in the current execution. Thus,  $e_i \in \mathcal{E}$ is the event which fired the transition from state $s_i$ to an initial state of another view state machine. In the following, $\epsilon$ represents the {\em empty} (event) history.

The evolution of configurations is given by the transition relation $\xrightarrow{-}_d$   defined by the rules in Figure~\ref{fig:formal:rules}. Relation $\xrightarrow{-}_d$ is constructed from the transition relations of view state machines $\xrightarrow{-}_i$, and the binary connection relation $\xrightarrow{-}_c$. In these rules, given a history of states $s_1\cdot \cdots \cdot s_n$ and the index $j$ of a view state machine $M_j$, function $top:\Sigma^* \times {\mathcal N} \to \Sigma \cup \{\bot\}$ returns the last state of the view state machine $M_j$ in the sequence $s_1 \cdot \cdots \cdot s_n$. That is, $top(s_1 \cdot \cdots \cdot s_n,j)$  returns  $s_k$, if $1 \le k \le n$ is the biggest index such that $s_k \in \Sigma_j$, or $\bot$, if such a state does not exist.

{
\begin{figure}
\begin{center}
\begin{tabular}{ll}

\textbf{R1.} ${\displaystyle\frac{s \xrightarrow{e}_i s' }{\langle s,h,eh \rangle \xrightarrow{e }_d \langle s',h,eh \rangle}}$ &
\textbf{R2.} ${\displaystyle\frac{s \in C_i, s \xrightarrow{e}_c s', \neg reuse(e)}{\langle s,h,eh \rangle \xrightarrow{e }_d \langle s',h\cdot return(s),eh\cdot e \rangle}}$ \\ [3ex]
\multicolumn{2}{l}{\textbf{R3.} ${\displaystyle \frac{s \in C_i, s'\in I_j, s \xrightarrow{e}_c s',  reuse(e), top(s_1 \cdots s_n,j) = \bot}{\langle s,h,eh \rangle \xrightarrow{e }_d \langle s',h\cdot return(s),eh\cdot e \rangle}}$} \\ [3ex]
\multicolumn{2}{l}{\textbf{R4.}  ${\displaystyle\frac{s \in C_i, s'\in I_j, s \xrightarrow{e}_c s', reuse(e), top(s_1\cdots s_n,j) = s_k }{\langle s,s_1 \cdots s_n,e_1\cdots e_n \rangle \xrightarrow{e }_d \langle s_k,s_1 \cdots  s_{k-1},e_1 \cdots e_{k-1} \rangle}}$} \\ [3ex]
\textbf{R5.}  ${\displaystyle\frac{s \in F_i, auto\_return(e)}{\langle s,h\cdot s',eh\cdot e \rangle \xrightarrow{-}_d \langle s',h,eh \rangle}}$ \\ [3ex]
\textbf{R6.}
\begin{math}
  \displaystyle\frac{c_0 \xrightarrow{e^+}_d c_1 }
  {\langle c_0,c_0',dh \rangle \xrightarrow{e^+}_{d||d'} \langle c_1,c_0',dh+\{e^+\} \rangle}
\end{math} &
\textbf{R7.}
\begin{math}
  \displaystyle\frac{c_0' \xrightarrow{e^-}_{d'} c_1',\ e^+ \in dh}
  {\langle c_0,c_0',dh \rangle \xrightarrow{e^-}_{d||d'} \langle c_0,c_1',dh-\{e^+\} \rangle}
\end{math} \\ [3ex]

\end{tabular}
\end{center}
\caption{Transition relation rules}
\label{fig:formal:rules}
\end{figure}
}

Rule {\bf R1} states that a transition inside a  view state machine $M_i$ corresponds to a transition in the device state machine. Rules {\bf R2, R3} model a transition from machine $M_i$ to machine $M_j$ when both the new state $s'$ and the event $e$ are added to the state and event histories of the current system configuration. Rule {\bf R2} is applied when event $e$ does not involve reusing a previously created view ($reuse(e)$ is false), while {\bf R3}  applies when a view of $M_j$, should have been reused ($reuse(e)$ is true), but the current state history does not contain one ($top(s_1 \cdots s_n,j) = \bot$). Rule {\bf R4} defines a transition from machine $M_i$ to $M_j$ by reusing  a previously created flow of $M_j$ ($reuse(e)$ is true) which is stored in the configuration history ($top(s_1\cdots s_n,j) = s_k$). Finally, {\bf R5} defines the case when the flow of the current active view has finished, and the execution must continue with the view stored at the top of the state history. Otherwise, that is, if  $auto\_return(e)$ returns false, the current configuration $\langle s,h,eh\rangle$ cannot evolve.

\begin{definition} Given a {\em device state machine}
\begin{align}
{\mathcal D} & = M_1|||\cdots|||M_n ||| \mathcal{R}(M_1,\cdots,M_n) \nonumber \\
& = \langle \Sigma \times \Sigma^*\times \mathcal{E}^*,I,\xrightarrow{-}_d,E \cup \mathcal{E},F \rangle \nonumber
\end{align}

\begin{enumerate}
\item the {\em trace-based semantics} determined by ${\mathcal D}$ (${\mathcal O}({\mathcal D})$) is given by the set of finite/infinite sequences of configurations (flows) produced by the transition relation  $\xrightarrow{-}_d$ from an initial state, that is, ${\mathcal O}({\mathcal D}) = \{\langle s_0,\epsilon,\epsilon\rangle \xrightarrow{e_0}_d \langle s_1,h_1,eh_1 \rangle \cdots | s_0 \in I\}$.

    \item Given a flow $\phi = c_0 \xrightarrow{e_1}_d c_1 \xrightarrow{e_2}_d c_2 \cdots \in {\mathcal O}({\mathcal D})$, the test case determined by $\phi$ is the sequence of events $test(\phi) = e_1 \cdot e_2 \cdots$
    \item The set of {\em test cases} determined by a set of flows ${\mathcal T}$  is  $TC({\mathcal T}) = \{test(t)| t \in {\mathcal T} \} $.
\end{enumerate}

\end{definition}

Thus, a flow $\phi \in {\mathcal O}({\mathcal D})$ consists of a sequence of view state machine flows (Definition~\ref{lab:flow-test-cases}) connected throw {\em connection states}. Flow $\phi$ may finish at a final state of some view state machine, or may be infinite.
 The {\em length} $|\phi|$ of a flow $\phi$ is the number of its states (configurations), if it is finite, or $\infty$, otherwise. Given a flow $\phi = c_0 \xrightarrow{e_1}_d c_1 \xrightarrow{e_2}_d c_2 \cdots \in {\mathcal O}({\mathcal D})$, we define the truncated flow of $n$, $\phi^n$, as $\phi$ iff $|\phi| <= n$ or $\phi^n = c_0 \xrightarrow{e_1}_d c_1 \xrightarrow{e_2}_d c_2 \cdots \xrightarrow{e_{n-1}}_d c_{n-1}$, otherwise.
 Considering this, we define the set of traces ${\mathcal O}^n({\mathcal D})$ as the set all traces of ${\mathcal O}({\mathcal D})$ truncated up to length $n$, that is, ${\mathcal O}^n({\mathcal D}) = \{  \phi^n | \phi \in {\mathcal O}({\mathcal D}) \}$.
 
Observe that the state space of device state machines is not finite because configurations include the state and event histories which may have arbitrary lengths. In addition, the state space generated when an explicit model checker is constructing all the flows allowed by a device state machine is  non-finite not only due to the state and event histories, but also because the matching algorithm, carried out during the state space search, must take into account both the current state of the flow and the history of the previous states of the flow. This allows that, for instance, flows $\phi_1=\langle s_0,\epsilon,\epsilon\rangle \xrightarrow{e_1^+}_d \langle s_1,\epsilon,\epsilon\rangle \xrightarrow{e_2^+}_d \langle s_2,\epsilon,\epsilon\rangle \xrightarrow{e_3^+}_d\langle s_3,\epsilon,\epsilon\rangle $ and
$\phi_2 = \langle s_0,\epsilon,\epsilon\rangle \xrightarrow{e_4^+}_d \langle s_4,\epsilon,\epsilon\rangle \xrightarrow{e_1^+}_d \langle s_1,\epsilon,\epsilon\rangle \xrightarrow{e_2^+}_d \langle s_2,\epsilon,\epsilon\rangle \xrightarrow{e_3^+}_d\langle s_3,\epsilon,\epsilon\rangle $ can be both generated by the model checker although when constructing $\phi_2$ state $s_1$ has been already visited  as explained in Section~\ref{sec:modelchecking}.
 
 In consequence, the models of device state machines are not, in general, state finite which means that, the model checking process does not, in general, terminate. In the current implementation, we have solved this problem by bounding the depth of the execution flows analyzed generating ${\mathcal O}^n({\mathcal D})$ for some fixed $n$.

\subsubsection{Composing several devices}
The extension of the state machine model to several devices is carried out by composing the device state machines by interleaving. Thus, if $c_0 \xrightarrow{e_1}_d c_1$  and $c'_0 \xrightarrow{e'_1}_{d'} c'_1$ are a transitions in devices ${\mathcal D} $ and ${\mathcal D}' $, respectively, then they allow the two  following transitions, $\langle c_0,c_0' \rangle \xrightarrow{e_1}_{d||d'} \langle c_1,c_0' \rangle$ and
$\langle c_0,c_0' \rangle \xrightarrow{e'_1}_{d||d'} \langle c_0,c_1'\rangle$ in the interleaved composition of ${\mathcal D} $  and ${\mathcal D'} $.
The communication between both devices is modeled by a user event in the sender device (the device that starts the communication), and a system event in the receiver device (the device that expects the message).

This is described in the last two rules of Figure~\ref{fig:formal:rules}. Rule \textbf{R6} handles the transition from the sender, and \textbf{R7} handles the transition in the receiver. Note that $dh$ denotes the set of system events produced but not yet consumed. Thus, for instance, using the previous example, if $e_1=e_1^+$ is an event that implies a communication from ${\mathcal D} $  to ${\mathcal D'}$, and $e'_1=e_1^-$ is the corresponding event to be read by ${\mathcal D'}$ from ${\mathcal D} $, we would generate the  test cases $e_1^+ \cdot e_1^-$ and
$e_1^- \cdot e_1^+$. Note that in the second test case, the method that implements the transition for the receiver event will suspend the execution of ${\mathcal D'}$  until event $e_1^+$ is fired by ${\mathcal D} $.

In addition, when dealing with more that one device, we make use of model checking optimization techniques such as partial order reduction~\cite{holzmann-03} to avoid the generation of different test cases that correspond to a single feasible interaction between the devices.

\section{Case Study}
\label{sec:Implementation}
\label{sec:casestudy}

In this section, we describe how the behavior of mobile applications is modeled and how tests cases are automatically generated from these models.

\subsection{Modeling}
\label{sec:statemachineabstraction}
\label{sec:casestudy:modeling}

We first need to construct an abstract model of the system to be analyzed, using statecharts and following the  notions of view and device state machines given in Section~\ref{sec:formal}. This model should include the relevant user interactions for the tests we want to perform. For instance, a test case which is affected by whether the GPS is on or off may include user interactions to change its status, while other tests may not need those interactions. A test case generator will be created from this model using automatic transformations. This modeling step can be performed separately from the design of the application, or in combination as is custom in other model-driven tools such as IBM Rational Rhapsody \cite{rhapsody-web}. In addition, the controls of each screen have to be extracted and modeled, so that transitions on the state machines can be tied to actions performed on these controls.

\begin{figure}[t]
\centering
\includegraphics[width=0.50\columnwidth]{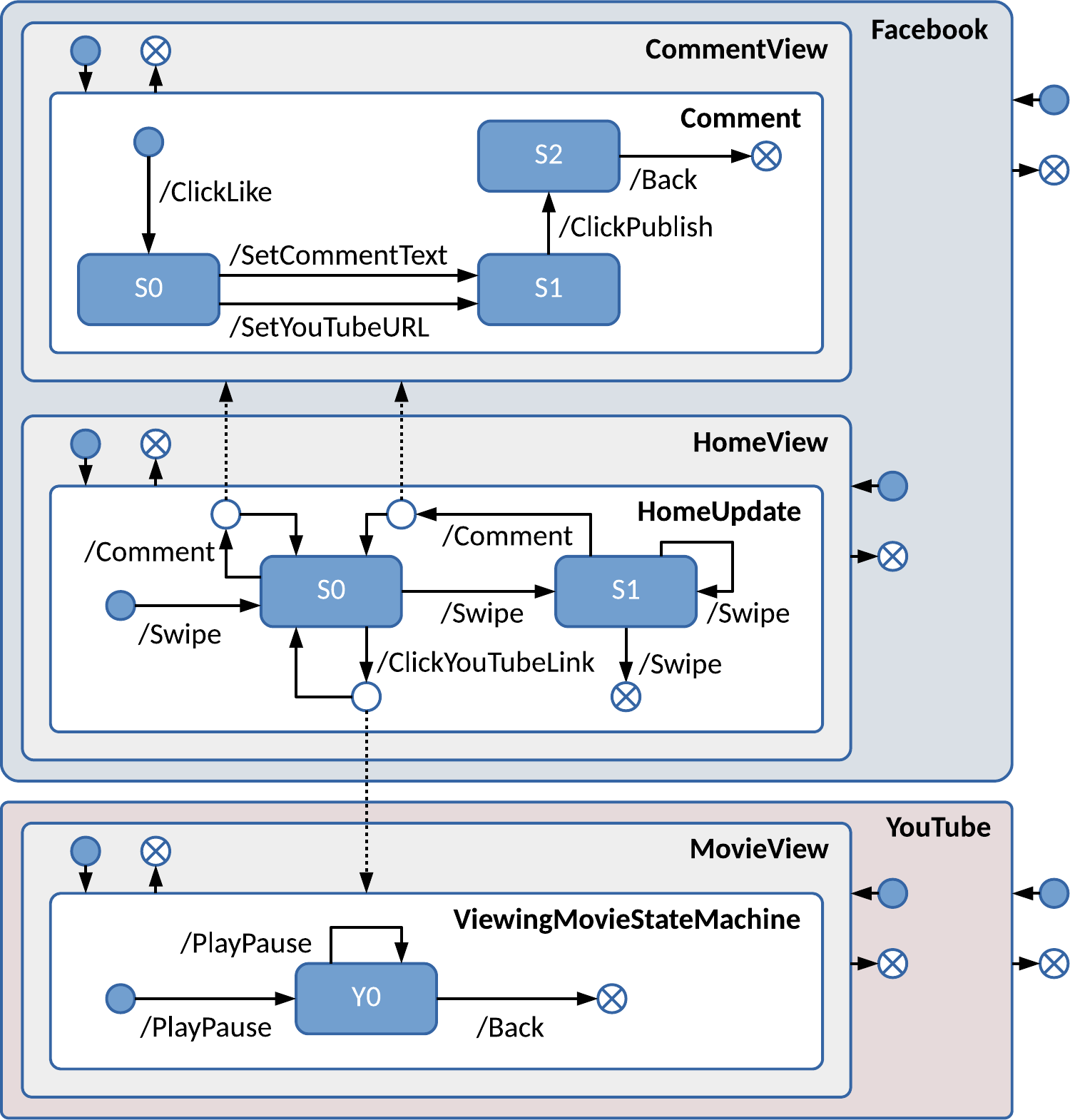}
\caption{Facebook and YouTube model}
    \label{fig:CompositionStateMachineYahoo}
\end{figure}

We use a scenario with two applications, \emph{Facebook} and \emph{YouTube}. This scenario is composed of three views (\emph{HomeView}, \emph{CommentView} and \emph{MovieView}), which describe the behavior of a user placing a link to a YouTube video in a Facebook comment, and watching this or other videos in  YouTube. The state machines can be modeled using UML  as shown in Figure~\ref{fig:CompositionStateMachineYahoo}. These state machines include additional information required to correlate them with the applications and their views. An \xml definition of the model can then be automatically generated from these state machines. Listing~\ref{lst:statemachine} shows part of this \xml definition~\footnote{More complete versions of this an other parts of the model are available online at \url{http://morse.uma.es}.}. In particular, it contains the state machine associated with the \emph{Home} view of the Facebook app. Each state machine may define several states and transitions. In addition to simple transitions between states of the same state machine, it is also possible to define transitions that call another state machine and, upon its termination, continue in the caller machine. The \texttt{type} and \texttt{through} attributes identify the type of transition and the state machine to call (if any). Listing~\ref{lst:statemachine} provides examples of both simple (line~\ref{lst:facebookyoutube:simpletransition}) and complex (lines~\ref{lst:facebookyoutube:viewtransition} and~\ref{lst:facebookyoutube:statemachinetransition}) transitions. Each transition has an unique \texttt{ID} within its view that is used to identify transitions, and also declares the user or system event that triggers it.

\begin{lstlisting}[style=xml,caption=State machine configuration,tabsize=2,float=t,label=lst:statemachine]
<Application name="Facebook" package="com.facebook.android">
	<Views>
		<View name="HomeView" controlsFile="Home.xml" >
			<StateMachines>
				<StateMachine name="HomeUpdate">
					<States><State name="S0"/><State name="S1"/></States>
					<Transitions>
						<Transition ID="1" event="Swipe" prev="" next="S0" type="Simple"/>
						/*@ \label{lst:facebookyoutube:viewtransition} @*/<Transition ID="2" event="Comment" prev="S0" next="S0" through="CommentView" type="View"/>
						/*@ \label{lst:facebookyoutube:simpletransition} @*/<Transition ID="3" event="Swipe" prev="S0" next="S1" type="Simple"/>
						/*@ \label{lst:facebookyoutube:statemachinetransition} @*/<Transition ID="4" event="ClickYouTubeLink" prev="S0" next="S0" through="ViewingMovieStateMachine" type="StateMachine"/>
						<Transition ID="5" event="Swipe" prev="S1" next="S1" type="Simple"/>
						<Transition ID="6" event="Comment" prev="S1" next="S0" through="CommentView" type="View"/>
						<Transition ID="7" event="Swipe" prev="S1" next="" type="Simple"/>
						...
\end{lstlisting}



\begin{figure}[t]
\centering
\includegraphics[width=0.5\columnwidth]{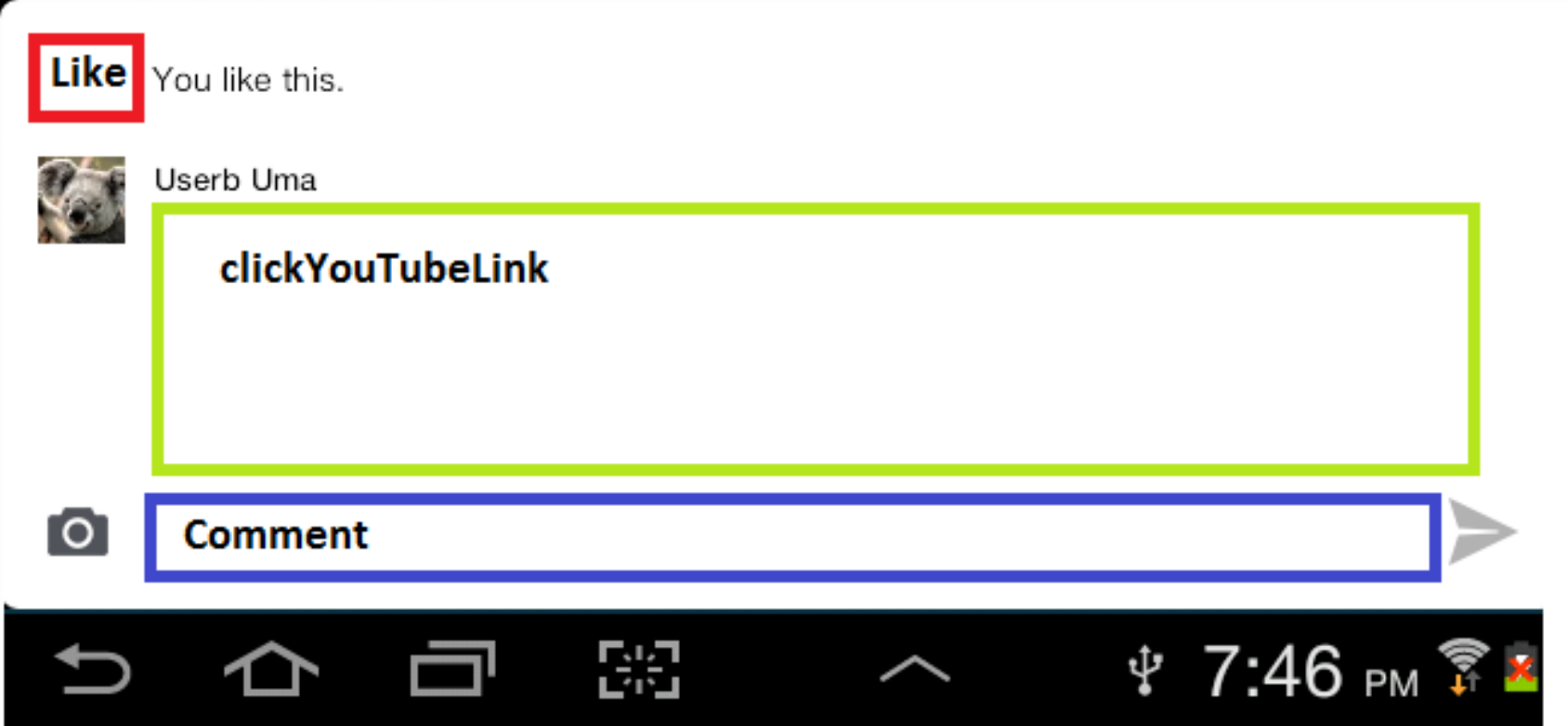}
\caption{Identifying control groups}
\label{fig:ControlDefinition}
\end{figure}

The events that fire the transitions in Figure~\ref{fig:CompositionStateMachineYahoo} are the user actions performed on controls placed in visible views. We organize controls into \emph{group of controls} according to the actions associated with each. Figure~\ref{fig:ControlDefinition} shows some of the control groups that have been identified in the \emph{CommentView} View. For instance, the \emph{Comment} group could represent any of the text fields to write a comment, and the \emph{clickYouTubeLink} identifies links to YouTube videos.

These control groups are declared in a \emph{control definition} file with the help of the \uiautomatorviewer tool~\cite{AndroidDev}. This tool analyzes each view without requiring its source code, and generates an UIX (\xml) file containing the hierarchy of controls in the view.
Listing~\ref{lst:controlgroups} shows part of the generated file for the \emph{Facebook} application. The attributes associated with each control in the UIX file include the kind of actions that the control supports, such as \texttt{clickable} or \texttt{scrollable}.
The UIX file is then customized to bring together the controls which belong to the same group by setting the \texttt{controlGroup} attribute. Some controls  accept parameters which may also be included as attributes in this file. For instance, the values introduced in text fields may be fixed or generated automatically according to some pattern.




\begin{lstlisting}[style=xml,caption=Control group definition,float=t,label=lst:controlgroups]
<node index="0" text="" testGroup="" ....
  <node index="0" ....
		<node testGroup="clicLike" IsFixedValue="" PatternOrValue="" index="0" text="Like" resource-id="id/feed_feedback_like_container" clickable="true" long-clickable="false" password="false" ... />
\end{lstlisting}

\subsection{Test case generation}
\label{sec:control}
\label{sec:casestudy:testcases}

We are now ready to generate the corresponding test cases in an exhaustive manner. The \xml file is automatically transformed into a \promela specification that follows the same principles described in Section~\ref{sec:modelchecking}, but with a few additions to acomodate the structure of \android applications. Each device is still represented by a single process, but their state machines are defined in separate \emph{inlines}, one per each app, view and state machine, which can then be composed. In addition, there may be device-specific app and view inlines, since views and state machines can be assigned to a particular device. In the simplest form of composition, device processes call their app inlines, app inlines call their view inlines, and view inlines call their state machine inlines. On the other hand, a state machine may call another view or state machine. When this happens, the state of the previous state machine should be stored such that when the new one is finished, the state is correctly restored. To support this we introduce a \emph{backstack} data structure, where the state of the current state machine is always at the top of the stack.

Listing~\ref{lst:control:testgeneration:promela} shows a simplified extract of the \promela specification generated for the
Facebook and YouTube example.
The backstack data structure is shown on line~\ref{lst:control:testgeneration:promela:backstack}. A new element is pushed to or popped from the backstack at the beginning or end of a state machine, respectively (lines~\ref{lst:control:testgeneration:promela:push} and~\ref{lst:control:testgeneration:promela:pop}), while \texttt{currentState} always point to the top of this stack for each device.

\begin{lstlisting}[caption={\promela specification for Facebook and YouTube},float=t,label=lst:control:testgeneration:promela]
typedef Backstack { mtype states[MAX_BK]; short index; }
Backstack backstacks[DEVICES]; /*@ \label{lst:control:testgeneration:promela:backstack} @*/
#define currentBackstack	devices[device].backstack
#define currentState		currentBackstack.states[currentBackstack.index]
active proctype device_219dcac41() { /*@ \label{lst:control:testgeneration:promela:deviceproca} @*/
	if
	true -> app_219dcac41_Facebook(D_219dcac41);
	true -> app_219dcac41_YouTube(D_219dcac41);
	fi;
	devices[D_219dcac41].finished = true
}
inline statemachine_Facebook_HomeView_HomeUpdate(device) {
	currentState = State_Facebook_HomeView_HomeUpdate_init;
	pushToBackstack(device, State_Facebook_HomeView_HomeUpdate_init); /*@ \label{lst:control:testgeneration:promela:push} @*/
	do
	:: currentState == State_Facebook_HomeView_HomeUpdate_S0 ->
		transition(device, VIEW_HomeView, 2);
		view_Facebook_CommentView(device); /*@ \label{lst:control:testgeneration:promela:callview} @*/
		currentState = State_Facebook_HomeView_HomeUpdate_S0
	:: currentState == State_Facebook_HomeView_HomeUpdate_S0 ->
		transition(device, VIEW_HomeView, 4);
		statemachine_YouTube_MovieView_ViewingMovieStateMachine(device); /*@ \label{lst:control:testgeneration:promela:callstatemachine} @*/
		currentState = State_Facebook_HomeView_HomeUpdate_S0
	...
	od;
	popFromBackstack(device) /*@ \label{lst:control:testgeneration:promela:pop} @*/
}
\end{lstlisting}

Each sequence of transitions generated by \spin is translated into a \texttt{UiAutomatorTestCase} subclass, where each transition is implemented by a method.
This class simulates the actions performed by the user, such as pressing buttons or swiping. The code shown in Listing~\ref{lst:uiautomatortestcase} shows part of a test case obtained from the model of Figure~\ref{fig:CompositionStateMachineYahoo}, in particular a user adding a link to a YouTube video in a Facebook comment, and later watching that video on the YouTube application.
The file is compiled into a \texttt{.dex} (\android application binary) file, and then deployed into a \android device and executed using the \emph{adb} tool.

\begin{lstlisting}[language=Java,caption=Generated UiAutomatorTestCase,float=t,label=lst:uiautomatortestcase]
public class TestDevice1 extends UiAutomatorTestCase {
	// Transition  2  previous S0 next S0 on view HomeView
	public void TestFacebookComment2() throws UiObjectNotFoundException {
		UiObject control = new UiObject(new UiSelector(). className("android.widget.TextView").index(1).textContains("Comment"));
		control.click();
	}
	// Transition  4:  previous S0 next S0 on view HomeView
	public void TestFacebookclicYouTubeLink27() throws UiObjectNotFoundException {
		UiObject control = new UiObject(new UiSelector().className("android.view.View").index(3));
		control.click();
	}
	// Transition  1:  previous  next Y0 on view MovieView
	public void TestYouTubeplaypause28() throws UiObjectNotFoundException {
		UiObject control = new UiObject(new UiSelector().className("android.view.View").index(4));
		control.click();
	}
}
\end{lstlisting}


%

Table~\ref{tab:casestudy:testcases:results} provides some quantitative results of the number of test cases generated and the computational effort required, for several scenarios, averaged over three runs. Device 219dcac4 was assigned only the Facebook application, while device 219dcac41 was assigned both, although in both cases other modeled applications may be reached from the assigned ones. The fourth column declares the maximum depth allowed for the test case transitions generated for a device. The fifth column represents the total time spent to generate the test cases from the \xml models. The last three columns are stats taken from the \spin execution, namely the number of \spin states generated, the size of each state, and the total memory spent, respectively. These results show how adding the YouTube application, which is fairly isolated, has little impact in the results (rows 3 and 4 of data).

{
\renewcommand{\tabcolsep}{4pt}

\begin{table*}[t]
\centering
\footnotesize
\begin{tabular}{|r|r|r|r|r|r|r|r|r|}
\hline
\multicolumn{2}{|c|}{Devices}	& \multicolumn{2}{|c|}{Configuration}	& \multicolumn{5}{|c|}{Results}	\\ \hline
219dcac4	& 219dcac41			& Backstack	& Transitions				& \# Test Cases	& Time (s)	& \# States	& State Size (B)	& Memory (MB) \\ \hline
\checkmark	&					& 4			& 20						& 5641			& 1.0		& 307234	& 84				& 156.8		\\
\checkmark	&					& 4			& 26						& 111317		& 9.0		& 6063398	& 92				& 728.6		\\
			& \checkmark		& 4			& 20						& 5660			& 1.0		& 307493	& 84				& 156.8		\\
			& \checkmark		& 4			& 26						& 111342		& 9.0		& 6063735	& 92				& 728.6		\\
\checkmark	& \checkmark		& 4			& 10						& 1872			& 7.0		& 4039337	& 100				& 560.3		\\
\checkmark	& \checkmark		& 4			& 12						& 12180			& 52.3		& 28972472	& 108				& 3445.2	\\
\hline
\end{tabular}
\caption{Test case generation results}
\label{tab:casestudy:testcases:results}
\end{table*}

}

\section{Comparison with related work}
\label{sec:related}

There are other proposals to apply model-based testing to \android applications. Some of them consider that the testing process starts without a precise model of the expected behavior of the applications and they focus on techniques to obtain such model.  {\em MobiGUITAR} framework \cite{AFT2014} automatically construct a state machine of one application by executing events in the running application and recording a tree with fireable events for each new state. The authors use a "breadth-first" traversal of the apps GUI for open source applications. As far as they are not considering any knowledge on the way of using  the application  but they are making an exhaustive  execution, they need some criteria to assume whether some states are equivalent to prevent state explosion.  The {\em Swift-Hand} technique proposed in \cite{CNS2013} employs machine learning to construct an approximated model of the application during the testing process. Their aim is to cover as much behavior as possible, making the execution to enter in unexplored parts of the state space.
In our method, we separate test generation from testing and the states in our high-level state machines are limited and differentiated by design. So our models are more compact, and for instance, compared with {\em MobiGUITAR}  we do not need extra work to remove unrealistic test cases. 
In addition, our approach allows to generate test cases for several applications that interact using \android\ intents, while the complexity of the runtime based modeling process for {\em MobiGUITAR} and {\em Swift-Hand} makes them more suitable for single applications.

Like in our proposal, other works also  consider the existence of a formal specification of the applications to start the test generation.In  \cite{JingAH12} the authors describe how to follow a property-driven method build the models in Alloy, a formal language based on high order logic. In their proposal the role of the model checker in our approach is done they  the Alloy analyzer, that generates positive (expected) and negative (undesired) test cases. Like in our approach, they use \xml based transformations to translate the test cases to some executable form in order to activate the applications under test. Apart from the inside technologies (model checking vs constraint solver), the main difference in both proposals is the way to obtain the refined executable model. The Alloy specification in \cite{JingAH12} is constructed manually, while the \promela specification in our work is done automatically from the high level design of the view state machines. We still need to work in the same case study to get a quantitative comparison on the human and computational effort required in both approaches. 
 
There are other model-based testing tools for Android which are not focused on models that consider the  user inputs. For instance, the tool APSET~\cite{APSET} considers manually constructed formal models of vulnerability patterns to generate test cases for \android\ applications. Test generation is implemented with an ad-hoc algorithm that also considers the compiled code of the application and the configuration files in the \android\ system.

\section{Conclusions and Future Work}
\label{sec:conclusiones}

\android systems have a complex architecture designed to support the concurrent execution of applications on devices with limited resources. Here we have presented a model-based testing approach for generating test cases for \android applications, which takes into account the way in which these applications interact with the user and with each other.
We model the expected user behavior by composing state machines, and then explore this model exhaustively with \spin to obtain all possible user behaviors, which correspond to test cases. These test cases are then executed in the device simulating the user inputs. In contrast with other approaches that generate random input events, our approach produces realistic user behaviors. Although our tool is currently geared towards \android, the same principles can be applied to analyze applications in other mobile platforms, such as {\sc iOS} and {\sc Windows Mobile}.

The next step of our work will be to connect the generated test cases with a runtime verification monitor \dragonfly~\cite{Drangofly,prole}.  In addition, we are working in adding more runtime information, like energy consumption, to perform richer analysis.

\bibliographystyle{eptcs}

\begin{thebibliography}{10}
\providecommand{\bibitemdeclare}[2]{}
\providecommand{\surnamestart}{}
\providecommand{\surnameend}{}
\providecommand{\urlprefix}{Available at }
\providecommand{\url}[1]{\texttt{#1}}
\providecommand{\href}[2]{\texttt{#2}}
\providecommand{\urlalt}[2]{\href{#1}{#2}}
\providecommand{\doi}[1]{doi:\urlalt{http://dx.doi.org/#1}{#1}}
\providecommand{\bibinfo}[2]{#2}

\bibitemdeclare{misc}{AndroidDev}
\bibitem{AndroidDev}
\emph{\bibinfo{title}{Android developers}}.
\newblock \bibinfo{note}{Http://developer.android.com/}.

\bibitemdeclare{misc}{DroidPilot}
\bibitem{DroidPilot}
\emph{\bibinfo{title}{{DroidPilot}}}.
\newblock \bibinfo{note}{Http://droidpilot.wordpress.com/}.

\bibitemdeclare{misc}{rhapsody-web}
\bibitem{rhapsody-web}
\emph{\bibinfo{title}{{IBM - Software - Rational Rhapsody family}}}.
\newblock \bibinfo{note}{Http://www-01.ibm.com/software/awdtools/rhapsody/}.

\bibitemdeclare{misc}{Robotium}
\bibitem{Robotium}
\emph{\bibinfo{title}{Robotium}}.
\newblock \bibinfo{note}{Https://code.google.com/p/robotium/}.

\bibitemdeclare{article}{AFT2014}
\bibitem{AFT2014}
\bibinfo{author}{Domenico \surnamestart Amalfitano\surnameend},
  \bibinfo{author}{Anna~Rita \surnamestart Fasolino\surnameend},
  \bibinfo{author}{Porfirio \surnamestart Tramontana\surnameend},
  \bibinfo{author}{Bryan \surnamestart Ta\surnameend} \& \bibinfo{author}{Atif
  \surnamestart Memon\surnameend} (\bibinfo{year}{2014}):
  \emph{\bibinfo{title}{{MobiGUITAR} -- A Tool for Automated Model-Based
  Testing of Mobile Apps}}.
\newblock {\sl \bibinfo{journal}{IEEE Software}}
  \bibinfo{volume}{99}(\bibinfo{number}{PrePrints}), p.~\bibinfo{pages}{1},
  \doi{10.1109/MS.2014.55}.

\bibitemdeclare{article}{ABGH2005}
\bibitem{ABGH2005}
\bibinfo{author}{Cyrille \surnamestart Artho\surnameend},
  \bibinfo{author}{Howard \surnamestart Barringer\surnameend},
  \bibinfo{author}{Allen \surnamestart Goldberg\surnameend},
  \bibinfo{author}{Klaus \surnamestart Havelund\surnameend},
  \bibinfo{author}{Sarfraz \surnamestart Khurshid\surnameend},
  \bibinfo{author}{Mike \surnamestart Lowry\surnameend},
  \bibinfo{author}{Corina \surnamestart Pasareanu\surnameend},
  \bibinfo{author}{Grigore \surnamestart Rosu\surnameend},
  \bibinfo{author}{Koushik \surnamestart Sen\surnameend},
  \bibinfo{author}{Willem \surnamestart Visser\surnameend} \&
  \bibinfo{author}{Rich \surnamestart Washington\surnameend}
  (\bibinfo{year}{2005}): \emph{\bibinfo{title}{Combining Test Case Generation
  and Runtime Verification}}.
\newblock {\sl \bibinfo{journal}{Theor. Comput. Sci.}}
  \bibinfo{volume}{336}(\bibinfo{number}{2-3}), pp. \bibinfo{pages}{209--234},
  \doi{10.1016/j.tcs.2004.11.007}.

\bibitemdeclare{article}{ArincTester}
\bibitem{ArincTester}
\bibinfo{author}{Pedro \surnamestart de~la C\'{a}mara\surnameend},
  \bibinfo{author}{J.~R\'{a}ul \surnamestart Castro\surnameend},
  \bibinfo{author}{Mar\'{\i}a \surnamestart del Mar~Gallardo\surnameend} \&
  \bibinfo{author}{Pedro \surnamestart Merino\surnameend}
  (\bibinfo{year}{2010}): \emph{\bibinfo{title}{Verification support for
  {ARINC}-653-based avionics software}}.
\newblock {\sl \bibinfo{journal}{Software Testing Verification \& Reliability}}
  \bibinfo{volume}{21}(\bibinfo{number}{4}), pp. \bibinfo{pages}{267--298},
  \doi{10.1002/stvr.422}.

\bibitemdeclare{article}{CNS2013}
\bibitem{CNS2013}
\bibinfo{author}{Wontae \surnamestart Choi\surnameend}, \bibinfo{author}{George
  \surnamestart Necula\surnameend} \& \bibinfo{author}{Koushik \surnamestart
  Sen\surnameend} (\bibinfo{year}{2013}): \emph{\bibinfo{title}{Guided GUI
  Testing of Android Apps with Minimal Restart and Approximate Learning}}.
\newblock {\sl \bibinfo{journal}{SIGPLAN Not.}}
  \bibinfo{volume}{48}(\bibinfo{number}{10}), pp. \bibinfo{pages}{623--640},
  \doi{10.1145/2544173.2509552}.

\bibitemdeclare{book}{CGP1999}
\bibitem{CGP1999}
\bibinfo{author}{Edmund~M. \surnamestart Clarke\surnameend, Jr.},
  \bibinfo{author}{Orna \surnamestart Grumberg\surnameend} \&
  \bibinfo{author}{Doron~A. \surnamestart Peled\surnameend}
  (\bibinfo{year}{1999}): \emph{\bibinfo{title}{Model Checking}}.
\newblock \bibinfo{publisher}{MIT Press}, \bibinfo{address}{Cambridge, USA}.

\bibitemdeclare{article}{SocketMC}
\bibitem{SocketMC}
\bibinfo{author}{Pedro \surnamestart de~la Cámara\surnameend},
  \bibinfo{author}{María \surnamestart del Mar~Gallardo\surnameend},
  \bibinfo{author}{Pedro \surnamestart Merino\surnameend} \&
  \bibinfo{author}{David \surnamestart Sanán\surnameend}
  (\bibinfo{year}{2009}): \emph{\bibinfo{title}{Checking the reliability of
  socket based communication software}}.
\newblock {\sl \bibinfo{journal}{Intl. Journal on Software Tools for Technology
  Transfer}} \bibinfo{volume}{11}(\bibinfo{number}{5}), pp.
  \bibinfo{pages}{359--374}, \doi{10.1007/s10009-009-0112-7}.

\bibitemdeclare{inproceedings}{Dwyer04}
\bibitem{Dwyer04}
\bibinfo{author}{M.B. \surnamestart Dwyer\surnameend},
  \bibinfo{author}{\surnamestart Robby\surnameend},
  \bibinfo{author}{O.~\surnamestart Tkachuk\surnameend} \&
  \bibinfo{author}{W.~\surnamestart Visser\surnameend} (\bibinfo{year}{2004}):
  \emph{\bibinfo{title}{Analyzing interaction orderings with model checking}}.
\newblock In: {\sl \bibinfo{booktitle}{Automated Software Engineering, 2004.
  Proceedings. 19th Intl. Conference on}}, pp. \bibinfo{pages}{154--163},
  \doi{10.1109/ASE.2004.1342733}.

\bibitemdeclare{inproceedings}{EGCC2010}
\bibitem{EGCC2010}
\bibinfo{author}{William \surnamestart Enck\surnameend}, \bibinfo{author}{Peter
  \surnamestart Gilbert\surnameend}, \bibinfo{author}{Byung-Gon \surnamestart
  Chun\surnameend}, \bibinfo{author}{Landon~P. \surnamestart Cox\surnameend},
  \bibinfo{author}{Jaeyeon \surnamestart Jung\surnameend},
  \bibinfo{author}{Patrick \surnamestart McDaniel\surnameend} \&
  \bibinfo{author}{Anmol~N. \surnamestart Sheth\surnameend}
  (\bibinfo{year}{2010}): \emph{\bibinfo{title}{TaintDroid: An Information-flow
  Tracking System for Realtime Privacy Monitoring on Smartphones}}.
\newblock In: {\sl \bibinfo{booktitle}{Proceedings of the 9th USENIX OSDI}},
  \bibinfo{series}{OSDI'10}, \bibinfo{publisher}{USENIX Association},
  \bibinfo{address}{Berkeley, CA, USA}, pp. \bibinfo{pages}{1--6},
  \doi{10.1145/2494522}.
\newblock \urlprefix\url{http://dl.acm.org/citation.cfm?id=1924943.1924971}.

\bibitemdeclare{inproceedings}{prole}
\bibitem{prole}
\bibinfo{author}{Ana~Rosario \surnamestart Espada\surnameend},
  \bibinfo{author}{María-del-Mar \surnamestart Gallardo\surnameend} \&
  \bibinfo{author}{Damián \surnamestart Adalid\surnameend}
  (\bibinfo{year}{2013}): \emph{\bibinfo{title}{DRAGONFLY : Encapsulating
  Android for Instrumentation}}.
\newblock In: {\sl \bibinfo{booktitle}{Proceedings of the XIII PROLE13}}.

\bibitemdeclare{inproceedings}{Drangofly}
\bibitem{Drangofly}
\bibinfo{author}{Ana~Rosario \surnamestart Espada\surnameend},
  \bibinfo{author}{María-del-Mar \surnamestart Gallardo\surnameend} \&
  \bibinfo{author}{Damián \surnamestart Adalid\surnameend}
  (\bibinfo{year}{2013}): \emph{\bibinfo{title}{A Runtime Verification
  Framework for android Applications}}.
\newblock In: {\sl \bibinfo{booktitle}{Proceedings of XXI JCSD}}.

\bibitemdeclare{book}{holzmann-03}
\bibitem{holzmann-03}
\bibinfo{author}{Gerard~J. \surnamestart Holzmann\surnameend}
  (\bibinfo{year}{2003}): \emph{\bibinfo{title}{{The {SPIN} Model Checker:
  Primer and Reference Manual}}}.
\newblock \bibinfo{publisher}{Addison-Wesley Professional}.

\bibitemdeclare{inproceedings}{Hu}
\bibitem{Hu}
\bibinfo{author}{Cuixiong \surnamestart Hu\surnameend} \&
  \bibinfo{author}{Iulian \surnamestart Neamtiu\surnameend}
  (\bibinfo{year}{2011}): \emph{\bibinfo{title}{Automating GUI Testing for
  Android Applications}}.
\newblock In: {\sl \bibinfo{booktitle}{Proceedings of the 6th International
  Workshop on AST}}, \bibinfo{series}{AST '11}, \bibinfo{publisher}{ACM},
  \bibinfo{address}{New York, NY, USA}, pp. \bibinfo{pages}{77--83},
  \doi{10.1145/1982595.1982612}.

\bibitemdeclare{inproceedings}{JingAH12}
\bibitem{JingAH12}
\bibinfo{author}{Yiming \surnamestart Jing\surnameend},
  \bibinfo{author}{Gail{-}Joon \surnamestart Ahn\surnameend} \&
  \bibinfo{author}{Hongxin \surnamestart Hu\surnameend} (\bibinfo{year}{2012}):
  \emph{\bibinfo{title}{Model-Based Conformance Testing for Android}}.
\newblock In: {\sl \bibinfo{booktitle}{Advances in Information and Computer
  Security - 7th International Workshop on Security, {IWSEC} 2012, Fukuoka,
  Japan, November 7-9, 2012. Proceedings}}, pp. \bibinfo{pages}{1--18},
  \doi{10.1007/978-3-642-34117-5\_1}.

\bibitemdeclare{inproceedings}{KFBJB2014}
\bibitem{KFBJB2014}
\bibinfo{author}{William \surnamestart Klieber\surnameend},
  \bibinfo{author}{Lori \surnamestart Flynn\surnameend}, \bibinfo{author}{Amar
  \surnamestart Bhosale\surnameend}, \bibinfo{author}{Limin \surnamestart
  Jia\surnameend} \& \bibinfo{author}{Lujo \surnamestart Bauer\surnameend}
  (\bibinfo{year}{2014}): \emph{\bibinfo{title}{Android Taint Flow Analysis for
  App Sets}}.
\newblock In: {\sl \bibinfo{booktitle}{Proceedings of the 3rd ACM SIGPLAN
  International Workshop}}, \bibinfo{series}{SOAP '14},
  \bibinfo{publisher}{ACM}, \bibinfo{address}{New York, NY, USA}, pp.
  \bibinfo{pages}{1--6}, \doi{10.1145/2614628.2614633}.

\bibitemdeclare{inproceedings}{LX2013}
\bibitem{LX2013}
\bibinfo{author}{Yepang \surnamestart Liu\surnameend} \& \bibinfo{author}{Chang
  \surnamestart Xu\surnameend} (\bibinfo{year}{2013}):
  \emph{\bibinfo{title}{VeriDroid: Automating Android application
  verification}}.
\newblock In: {\sl \bibinfo{booktitle}{Proceedings Middleware 2013 Doctoral
  Symposium}}, \bibinfo{publisher}{ACM}, \doi{10.1145/2541534.2541594}.

\bibitemdeclare{inproceedings}{MTN2013}
\bibitem{MTN2013}
\bibinfo{author}{Aravind \surnamestart Machiry\surnameend},
  \bibinfo{author}{Rohan \surnamestart Tahiliani\surnameend} \&
  \bibinfo{author}{Mayur \surnamestart Naik\surnameend} (\bibinfo{year}{2013}):
  \emph{\bibinfo{title}{Dynodroid: An Input Generation System for Android
  Apps}}.
\newblock In: {\sl \bibinfo{booktitle}{Proceedings of the 2013 ESEC/FSE}},
  \bibinfo{series}{ESEC/FSE 2013}, \bibinfo{publisher}{ACM},
  \bibinfo{address}{New York, NY, USA}, pp. \bibinfo{pages}{224--234},
  \doi{10.1145/2491411.2491450}.

\bibitemdeclare{article}{jpf2}
\bibitem{jpf2}
\bibinfo{author}{Peter \surnamestart Mehlitz\surnameend},
  \bibinfo{author}{Oksana \surnamestart Tkachuk\surnameend} \&
  \bibinfo{author}{Mateusz \surnamestart Ujma\surnameend}
  (\bibinfo{year}{2011}): \emph{\bibinfo{title}{JPF-AWT: Model checking GUI
  applications}}.
\newblock {\sl \bibinfo{journal}{2011 26th IEEE/ACM International Conference
  ASE 2011}} \bibinfo{volume}{0}, pp. \bibinfo{pages}{584--587},
  \doi{10.1109/ASE.2011.6100131}.

\bibitemdeclare{article}{jpf}
\bibitem{jpf}
\bibinfo{author}{Heila \surnamestart van~der Merwe\surnameend},
  \bibinfo{author}{Brink \surnamestart van~der Merwe\surnameend} \&
  \bibinfo{author}{Willem \surnamestart Visser\surnameend}
  (\bibinfo{year}{2012}): \emph{\bibinfo{title}{Verifying Android Applications
  Using Java PathFinder}}.
\newblock {\sl \bibinfo{journal}{SIGSOFT Softw. Eng. Notes}}
  \bibinfo{volume}{37}(\bibinfo{number}{6}), pp. \bibinfo{pages}{1--5},
  \doi{10.1145/2382756.2382797}.

\bibitemdeclare{article}{APSET}
\bibitem{APSET}
\bibinfo{author}{Sébastien \surnamestart Salva\surnameend} \&
  \bibinfo{author}{StassiaR. \surnamestart Zafimiharisoa\surnameend}
  (\bibinfo{year}{2014}): \emph{\bibinfo{title}{APSET, an Android aPplication
  SEcurity Testing tool for detecting intent-based vulnerabilities}}.
\newblock {\sl \bibinfo{journal}{International Journal on Software Tools for
  Technology Transfer}}, pp. \bibinfo{pages}{1--21},
  \doi{10.1007/s10009-014-0303-8}.

\bibitemdeclare{inproceedings}{TKH2011}
\bibitem{TKH2011}
\bibinfo{author}{Tommi \surnamestart Takala\surnameend}, \bibinfo{author}{Mika
  \surnamestart Katara\surnameend} \& \bibinfo{author}{Julian \surnamestart
  Harty\surnameend} (\bibinfo{year}{2011}): \emph{\bibinfo{title}{Experiences
  of System-Level Model-Based GUI Testing of an Android Application}}.
\newblock In: {\sl \bibinfo{booktitle}{Proceedings of the 2011 Fourth IEEE
  ICST}}, \bibinfo{series}{ICST '11}, \bibinfo{publisher}{IEEE Computer
  Society}, \bibinfo{address}{Washington, DC, USA}, pp.
  \bibinfo{pages}{377--386}, \doi{10.1109/ICST.2011.11}.

\bibitemdeclare{book}{UL2006}
\bibitem{UL2006}
\bibinfo{author}{Mark \surnamestart Utting\surnameend} \&
  \bibinfo{author}{Bruno \surnamestart Legeard\surnameend}
  (\bibinfo{year}{2007}): \emph{\bibinfo{title}{Practical Model-Based Testing:
  A Tools Approach}}.
\newblock \bibinfo{publisher}{Morgan Kaufmann Publishers Inc.},
  \bibinfo{address}{San Francisco, CA, USA}.

\bibitemdeclare{inproceedings}{YCZJ2013}
\bibitem{YCZJ2013}
\bibinfo{author}{Hui \surnamestart Ye\surnameend}, \bibinfo{author}{Shaoyin
  \surnamestart Cheng\surnameend}, \bibinfo{author}{Lanbo \surnamestart
  Zhang\surnameend} \& \bibinfo{author}{Fan \surnamestart Jiang\surnameend}
  (\bibinfo{year}{2013}): \emph{\bibinfo{title}{DroidFuzzer: Fuzzing the
  Android Apps with Intent-Filter Tag}}.
\newblock In: {\sl \bibinfo{booktitle}{Proceedings 11th International
  Conference on Advances in MoMM2013}}, \bibinfo{publisher}{ACM},
  \doi{10.1145/2536853.2536881}.

\end{thebibliography}

%

\end{document}